\documentclass[a4paper,UKenglish, autoref, thm-restate]{lipics-v2019}

\usepackage[ruled]{algorithm2e}
\usepackage[nameinlink]{cleveref}
\hypersetup{colorlinks=true}

\usepackage{easy-todo}

\usepackage{booktabs}
\usepackage{subcaption}

\usepackage{tikz}
\usetikzlibrary{positioning,backgrounds,shadings}

\usepackage{listings}

\title{Modern Analysis of Sparse Random Projections} %TODO Please add

\titlerunning{Modern Analysis of Sparse Random Projections} %TODO optional, please use if title is longer than one line

\author{Maciej Skorski}{University of Luxembourg}{}{}{}%TODO mandatory, please use full name; only 1 author per \author macro; first two parameters are mandatory, other parameters can be empty. Please provide at least the name of the affiliation and the country. The full address is optional

\authorrunning{M. Skorski} %TODO mandatory. First: Use abbreviated first/middle names. Second (only in severe cases): Use first author plus 'et al.'

\Copyright{Maciej Skorski} %TODO mandatory, please use full first names. LIPIcs license is "CC-BY";  http://creativecommons.org/licenses/by/3.0/

 %TODO mandatory; please add comma-separated list of keywords

\category{} %optional, e.g. invited paper

\relatedversion{} %optional, e.g. full version hosted on arXiv, HAL, or other respository/website
\supplement{}%optional, e.g. related research data, source code, ... hosted on a repository like zenodo, figshare, GitHub, ...

\nolinenumbers %uncomment to disable line numbering

\EventEditors{John Q. Open and Joan R. Access}
\EventNoEds{2}
\EventLongTitle{ }
\EventShortTitle{ }
\EventAcronym{ }
\EventYear{ }
\EventDate{December 24--27, 2016}
\EventLocation{Little Whinging, United Kingdom}
\EventLogo{}
\SeriesVolume{ }
\ArticleNo{ }
\DeclareMathOperator{\Even}{Even}

\begin{document}

\maketitle

%TODO mandatory: add short abstract of the document
\begin{abstract}
There has been recently a lot of research on sparse variants of random projections, faster adaptations of
the state-of-the-art dimensionality reduction technique originally due to Johsnon and Lindenstrauss. Although the construction is very simple, its analyses are notoriously complicated. Meeting the demand for both simplicity and accuracy, this work establishes sharp sub-poissonian tail bounds for the distribution of sparse random projections. Compared to other works, this analysis provide superior numerical guarantees (exactly matching impossibility results) while being arguably less complicated (the technique resembles Bennet's Inequality and is of independent interest).  
\end{abstract}

\section{Introduction}

\subsection{Contribution}

Random linear projections have the extremely attractive property of considerably \emph{reducing dimensionality}
of data while being \emph{nearly isometric}. This fact together with provable guarantees, shown for the first time by Johnson and Lindenstrauss~\cite{joag1983negative}, make them the state-of-the-art preprocessing technique for data science tasks and the focus of intensive research in high-dimensional statistics. More recently, there has been lots of works studying \emph{Sparse Johnson-Lindestrauss Transforms}
~\cite{dasgupta2010sparse,journals/eccc/KaneN10,braverman2010rademacher,kane2012sparser,cohen2016nearly,cohen2018simple}, in order to speed up computations. 

The purpose of this work is to give a novel analysis of Sparse Johnson-Lindestrauss Transforms, which is \emph{simple and accurate} at the same time. To this end, we re-analyze the sparse construction due to Kane~\cite{kane2012sparser}, which builds on randomly sparsifying Rademacher matrices offers best known dimension-quality tradeoffs. Our main result reads as follows: 

\begin{theorem}[Poisson Tails of Sparse Johnson-Lindenstrauss Lemma]\label{thm:main}
For any confidence $\delta >0$, distortion $\epsilon>0$ and integer sparsity $s>0$, there is a random $m\times n$ matrix $A$ with $p=\frac{s}{m}$ fraction of non-zero elements in each row, such that the nearly isometric property:
\begin{align}\label{eq:high_prob_JL}
     \mathbf{P}\left\{(1-\epsilon)\|x\|^2\leqslant \|Ax\|^2\leqslant (1+\epsilon)\|x\|^2\right\} \geqslant 1-\delta,
\end{align}
holds for any vector $x$, provided that the embedding dimension is at least
\begin{align}
    m \geqslant \frac{4\log(2/\delta)}{\epsilon^2}\cdot h\left(\frac{25\epsilon}{p}\right)^{-1},
\end{align}
and that $p\leqslant \frac{1}{30},\epsilon\leqslant p\log(1/2p)$, where $h$ is the "Bennet function" defined as 
\begin{align}
     h(u)\triangleq \frac{(1+u)\log(1+u)-u}{u^2/2}.
\end{align}
The construction of the matrix $A$ is presented in \Cref{alg:main}.
\end{theorem}

\begin{algorithm}[H]
\caption{Construction of Sparse Johnson-Lindenstrauss Transform}\label{alg:main}
\SetKwProg{SparseJLT}{SparseJLT}{}{}

\SparseJLT{$(n,m,p)$}{
\SetKwInOut{Input}{inputs}
 \SetKwInOut{Output}{output}
 \Input{data dimension $n$, embedding dimension $m$, sparsity $s$.}
 \Output{sparse projection $m\times n$ matrix $A$.}
\ForEach{column $i\in \{1\ldots m\}$}{
$A_i \gets $ sparse column of length $n$ with non-zero entries at $s$ random positions and values randomly chosen from
 $\pm \frac{1}{\sqrt{s}}$.
}
\KwRet{$A=[A_1\ldots A_m]$}
}
\end{algorithm}

The property \eqref{eq:high_prob_JL} is the standard formulation of the Johnson-Lindenstrauss Lemma (sometimes the result is presented as a union bound over multiple vectors $x$). Intuitively, it ensures the small relative change in the vector length (controlled by $\epsilon$) with high confidence ($\delta$ is the failure probability). The sparsity parameter $p$ is ideally kept possibly small, however the price is a loss in quality (bigger distortion, higher dimension or lower confidence). 

\subsection{Comparison to Related Works}
We now discuss how our result improves upon bounds available in the existing literature. 
\begin{enumerate}
    \item The result offers a valuable and elegant explanation of the tails in the sparse Johnson-Lindenstrauss Lemma: they are dominated by a Poisson distribution (the term $h$ is typical for Bennet's trick and Poissonian tail bounds). The proof is elementary and short.
    \item In the regime $\epsilon/p\to 0$, the bound exactly matches the best possible dimension for the Johnson-Lindenstrauss Lemma, found recently by Burr at al.~\cite{burr2018optimal} (we then have $m =  \frac{4\log(2/\delta)}{\epsilon^2}\cdot (1+o(1))$). This is a surprisingly strong result, as before we only knew that it approaches the optimal dimension up to a large constant factor (and the exponentially large gap loss in the confidence). In other words, this paper shows for the first time that sparse random projections can be on pair with their dense (unrestricted) counterparts.
    \item In the regime $\epsilon = \Theta(p)$ our bound shows that it is possible to choose $m=O(\frac{\log(2/\delta)}{\epsilon^2})$, which gives the sparsity $s=O(\epsilon^{-1})$. This is consistent with the literature, but our analysis gives much better constants (best explicit constants were due to Cohen at. al.~\cite{cohen2018simple})
    \item In the regime $\epsilon/p \gg 0$ our bound offers an interesting dimension-sparsity tradeoff. Namely, 
    we can have $m = \frac{4 B\log^{-1} B \log(2/\delta)}{\epsilon^2}$ and $s=O(\epsilon^{-1}/\log B)$. In other words, we can improve the sparsity by a factor, but loose exponentially in the dimension. Such a result appears only in one work, a somewhat complicated analysis due to Cohen~\cite{cohen2016nearly}. The result above is more elementary and provides explicit and better constants. It also appears to be asymptotically better, namely in ~\cite{cohen2016nearly} we find a bound on $m$ worse by $\log B$.
\end{enumerate}
Bounds from related works, (also the dimension optimallity), are covered in detail in \Cref{tab:summary}. 

%Clearly, our proof is superior in terms of the numerical constants. Furthermore, unlike other works, our bound shows that sparse random projections \emph{achieve the optimal dimension} in the regime $p=o(\epsilon)$ and $\epsilon=o(1),\delta=o(1)$. Finally, our proof is arguably simpler than the arguments given before.

\begin{table}[h]
    \centering
\resizebox{0.99\linewidth}{!}{
    \begin{tabular}{cll}
    \toprule
    Author & Bounds & Analysis \\
    \midrule
     \cite{kane2011almost} & $m \leqslant O\left(\frac{4\log(2/\delta)}{\epsilon^{-2}}\right)$ & Lower Bounds \\
     \cite{burr2018optimal} & $m \leqslant \frac{4\log(2/\delta)(1+o(1))}{\epsilon^{-2}},\ \epsilon\to 0$ & Lower Bounds \\
    \midrule
     \cite{journals/eccc/KaneN10} & $m \geqslant \Omega(1)\max\left\{ \frac{\log(2/\delta)}{\epsilon^2},\frac{\log^2(2/\delta)}{p \epsilon}\right\}$ &  Use of Rademacher Chaos Bounds \\
     \cite{journals/eccc/KaneN10} & $m \geqslant \Omega(1)\max\left\{ \frac{\log(2/\delta)}{\epsilon^2},\frac{\log^2(1/\delta)\frac{\log\log\log(2/\delta)}{\log\log(2/\delta)}}{p \epsilon}\right\}$ &  Moment Bounds by Graphs Enumeration \\
     \cite{kane2012sparser} & $m \geqslant \Omega(1)\max\left\{ \frac{\log(2/\delta)}{\epsilon^2},\frac{\log(2/\delta)}{p \epsilon}\right\}$ & Moment Bounds by Graphs Enumeration \\
         \cite{cohen2016nearly} & $m \geqslant \Omega(1)\max\left\{ \frac{B\log(2/\delta)}{\epsilon^2},\frac{\log_B(2/\delta)}{p \epsilon}\right\}, \textrm{any }B>2$ & Matrix Chernoff Bounds + Majorization Arguments\\
     \cite{cohen2018simple} & $m \geqslant \Omega(1)\max\left\{ \frac{\log(2/\delta)}{\epsilon^2}, \frac{\log(2/\delta)}{p \epsilon} \right\}$ & Hanson-Wright Lemma + Random Matrix Bounds \\
     \cite{cohen2018simple} & $m \geqslant \max\left\{ \frac{128\log(1/\delta)}{\epsilon^2}, \frac{8\sqrt{2}\log(2/\delta)}{p \epsilon} \right\}$ & Decoupling + Bounding MGF \\
    \midrule
     \textbf{this work} & $m\geqslant \frac{4\log(2/\delta)}{\epsilon^2}\cdot h\left(\frac{25\epsilon}{p}\right)^{-1},\quad \epsilon\leqslant p\log(1/2p)$ &  Bennet's Technique \\
    \bottomrule
    \end{tabular}
}
    \caption{Bounds for Sparse JL Transforms. It is assumed that $p$ is sufficiently small, e.g. $p<\frac{1}{30}$.}
    \label{tab:summary}
\end{table}

% The lower the matrix sparsity $p$, the worse the bound.

\section{Preliminaries}

The standard way of obtaining tail bounds is the Crammer-Chernoff inequality $\mathbf{P}\{X\geqslant \epsilon\}\leqslant \mathrm{e}^{-s\epsilon}\mathbf{E}\mathrm{e}^{sX}$, to be optimized over $s>0$ (Markov's inequality applied to exponential moments).

Recall that the Poisson random variable follows the distribution $\mathbf{P}\{\mathsf{Poiss}(\lambda)=k\}=\mathrm{e}^{-\lambda}\lambda^k/k!$. We have
$\mathbf{E}\mathrm{e}^{s\mathsf{Poiss}(\lambda)}=\mathrm{e}^{\lambda(\mathrm{e}^{t}-1)}$ and the Crammer-Chernoff method gives
\begin{lemma}[\cite{pollard2015mini}]\label{lemma:poiss}
For any $\lambda >0$ and $\epsilon>0$ we have that
 $\mathbf{P}\left\{\mathsf{Poiss}(\lambda)\geqslant \epsilon\right\}\leqslant \mathrm{e}^{-\lambda}(\mathrm{e}\lambda/\epsilon)^\epsilon$.
\end{lemma}

\section{Proof of \Cref{thm:main}}

\subsection{Step 1: Decomposition over Embedding Dimension}

Let $A$ be as in \Cref{alg:main}. By the identity $\|y\|^2 = y^Ty$ we obtain
\begin{align}
    \|A x\|^2-\|x\|^2 =  x^T B x,\quad B\triangleq A^T A - I.
\end{align}
We see that $B$ is off-diagonal, because $\sum_{k}A_{i,k}^2=1$ ($A$ has exactly $s$ entries of absolute value $1/\sqrt{s}$ in each column). Let $\eta_{i,j}$ indicate if the row $i$ is selected when constructing the column $j$ in 
\Cref{alg:main} and let $r_{i,j}$ be independent Rademacher random variables. Then:
\begin{align}
 B=\sum_k B_k,\quad  (B_k)_{i,j} = \begin{cases}
 \frac{1}{s} \eta_{k,i}\eta_{k,j}r_{k,i} r_{k,j} & i\not=j \\
 0 & i =j
\end{cases}.
\end{align}
Observe that $B_k$ are not independent, because $\eta_{k,i}$ are not independent across rows $k$. Nevertheless, one can show the following \emph{iid majorization}:
\begin{align}\label{eq:majorize}
0\leqslant \mathbf{E}(  x^T \sum_k B_k x)^q \leqslant \mathbf{E}(  x^T \sum_k B'_k x)^q,
\end{align}
where $(B'_k)$ are independent and distributed as $(B_k)$. Indeed,  $B'_k$ is be constructed by replacing $\eta_{k,i}$ that appear in $B_k$ by independent Bernoulli random variables $\eta'_{k,i}$. By expanding both sides into sums of monomials and the symmetry of $r_{k,i}$ we see that it suffices to prove
\begin{align}
    0 \leqslant \mathbf{E}\prod_{(k,i)\in S} \eta_{k,i}^{d_{k,i}} x_i^{d_{k,i}}  \leqslant 
   \mathbf{E} \prod_{(k,i)\in S} {\eta'}_{k,i}^{d_{k,i}} x_i^{d_{k,i}},
\end{align}
for any set of tuples $S$ and even exponents $d_{k,i}$. Now the first inequality follows as $d_{k,i}$ are even, while
the second reduces to $\mathbf{E}\prod_{i\in I} \eta_{k,i} \leqslant \mathbf{E}\prod_{i\in I} {\eta}'_{k,i}$ for any fixed $k$ and any set $I$ (because $\eta'_{k,i}$ are independent for different $k$), which simply means that sampling with replacement makes hitting any set $I$ less likely than independent sampling - obviously true.

We now see that the problem reduces to analyzing 
\begin{align}
    Z = \sum_{i\not=j} x_i x_j \eta_i \eta_j r_i r_j,
\end{align}
because from \Cref{eq:majorize} we have the following bound on exponential moments:
\begin{align}\label{eq:MGF_bound}
    \mathbf{E}\mathrm{e}^{t x^T B x} \leqslant \mathbf{E}\mathrm{e}^{|t| \sum_k x^T {B'}_k x} = \mathbf{E}\mathrm{e}^{t\cdot \frac{Z_1+\cdots + Z_m}{s}},\quad Z_i\sim^{iid} Z.
\end{align}

\subsection{Step 2: Moments of 1-Dim Dense Random Projections}

We first observe that for any sequence of real numbers $(y_i)$ and any positive integer $q$
\begin{align}
    (\sum_{i\not=j} y_i y_j)^q \leqslant (\sum_{i}y_i)^{2q}.
\end{align}
Indeed, if we replace $y_i$ by $|y_i|$ the left side can only increase while the right-hand side doesn't change; for non-negative $y_i$ the inequality follows as $\sum_{i\not=i}y_i y_j \leqslant (\sum_i y_i)^2$. Hence,
\begin{align}
    \mathbf{E}[Z^q] \leqslant \mathbf{E}(\sum_i x_i  \eta_i r_i)^{2q}.
\end{align}
By the multinomial expansion and the symmetry
\begin{align}
 \mathbf{E}(\sum_i x_i^2 \eta_i)^{2q} =    \sum_{r=2}^{q}p^r\sum_{d_1,\ldots,d_r\in\mathbb{Z}^{+}}\binom{2q}{2d_1,\ldots,2d_r}\sum_{i_1\not=i_2\not=\cdots\not=i_r}x_{i_1}^{2d_1}\cdots x_{i_r}^{2d_r}.
\end{align}
Furthermore, for every $d_1,\ldots,d_r\in\mathbb{Z}^{+}$ such that $\sum_i d_i = q$ it follows that
\begin{align}
    \sum_{i_1\not=i_2\not=\cdots\not=i_r}x_{i_1}^{2d_1}\cdots x_{i_r}^{2d_r} \leqslant \binom{q}{d_1,\ldots,d_r}^{-1} (\sum_{i} x_i^2)^{q},
\end{align}
by keeping only the terms $x_{i_1}^{2d_1}\cdots x_{i_r}^{2d_r}$ from the expansion of the right side.

By combining the last three inequalities and $\sum_i x_i^2=1$ we obtain
\begin{align}
    \mathbf{E} [Z^q] \leqslant \sum_{r=2}^{q}p^r\sum_{d_1,\ldots,d_r\in\mathbb{Z}^{+}}\binom{2q}{2d_1,\ldots,2d_r}\binom{q}{d_1,\ldots,d_r}^{-1}.
\end{align}
To simplify the sum, we will use the following bound
\begin{align}
    \binom{2q}{2d_1,\ldots,2d_r}\leqslant 2^q \cdot \binom{q}{d_1,\ldots,d_r}^2,\quad d_1,\ldots,d_r\in\mathbb{Z}^{+},d_1+\cdots+d_r=q.
\end{align}
The inequality is equivalent to $\binom{2q}{q}\leqslant 2^q\prod_{i=1}^{r}\binom{2d_i}{d_i}$.
Let $B_k = \binom{2k}{k}$ be the central binomial coefficient; the relation $B_{k}/B_{k-1} =4\cdot (1-1/2k)$ and the bound $1-\frac{1}{2k}\geqslant \frac{1}{2}$ imply $B_k \geqslant 2^k$. Then $2^q\prod\binom{2d_i}{d_i}\geqslant 2^q \cdot 2^{d_1+\cdots+d_r}=2^{2q}\geqslant \binom{2q}{q}$, so the claim follows.

%Using the simple bound $\binom{2d_i}{d_i}\geqslant 2^{2d_i}/2d_i$ (this follows as the central binomial coefficient is the biggest in the row of the Pascal triangle), we obtain $\prod_{i=1}^{r}\binom{2d_i}{d_i}\geqslant $

This bound, together with $\sum_{d_1,\ldots,d_r\in\mathbb{Z}^{+}}\binom{q}{d_1,\ldots,d_r}\leqslant (1+\cdots+1)^q=r^q$ implies
\begin{align}\label{eq:raw_moment_bound}
    \mathbf{E} [Z^q] \leqslant 2^{q}
    \sum_{r=2}^{q} p^r r^q.
\end{align}

\subsection{Step 3: Poisson Majorization}

Using \Cref{eq:raw_moment_bound} for $q\geqslant 3$ and the more accurate bound $\mathbf{E}[Z^q]\leqslant 2p^2$ for $q=2$,
we obtain the following bound on exponential moments (recall that $\mathbf{E}Z=0$):
\begin{align}
\begin{aligned}
    \mathbf{E}\mathrm{e}^{tZ}  &\leqslant 1+t^2p^2+\sum_{r\geqslant 2}p^r\sum_{q\geqslant \max\{3,r\}}\frac{(2tr)^q}{q!}  \\
    &=
    1+t^2p^2+\sum_{r\geqslant 2}p^r\mathrm{e}^{2tr}\mathbf{P}\left\{\mathsf{Poiss}(2tr)\geqslant \max\{3,r\}\right\}\\
    &=1+t^2p^2+p^2(\mathrm{e}^{4t}-8t^2-4t-1)+\sum_{r\geqslant 3}p^r\mathrm{e}^{2tr}\mathbf{P}\left\{\mathsf{Poiss}(2tr)\geqslant r\right\}
\end{aligned}.
\end{align}
Using the known bound $\mathbf{P}\left\{\mathsf{Poiss}(2tr)\geqslant r\right\}\leqslant \mathrm{e}^{-2tr}(2\mathrm{e}t)^r$ for the Poisson distribution tails (it is obtained from the moment generating function, see~\cite{pollard2015mini} or the next section for an implicit derivation) for $t<\frac{1}{2}$ and $\mathbf{P}\left\{\mathsf{Poiss}(2tr)\geqslant r\right\}\leqslant 1$ for $t\geqslant \frac{1}{2}$
we obtain the following bound
\begin{align}
    \mathbf{E}\mathrm{e}^{tZ} \leqslant 1+t^2p^2+p^2(\mathrm{e}^{4t}-8t^2-4t-1)+
\begin{cases}
        \frac{(2\mathrm{e}pt)^3}{1-2\mathrm{e}pt} & \quad 0<t<\frac{1}{2} \\
        \frac{p^3\mathrm{e}^{6t}}{1-p\mathrm{e}^{2t}} & \quad \frac{1}{2}\leqslant t<\frac{\log(1/p)}{2}
\end{cases}.
\end{align}
We transform the bound into the form

\newcommand{\parta}{{\frac{8\mathrm{e}^3 pt^3}{1-2\mathrm{e}pt}}}
\newcommand{\partb}{\frac{p\mathrm{e}^{6t}}{1-p\mathrm{e}^{2t}}}
\begin{align}
     \mathbf{E}\mathrm{e}^{tZ} \leqslant 1+t^2p^2 + p^2\cdot 
     \underbrace{
     \begin{cases}
     \mathrm{e}^{4t}-8t^2-4t-1 +
        \parta & \quad 0<t<\frac{1}{2} \\
    \mathrm{e}^{4t}-8t^2-4t-1 +    \partb & \quad \frac{1}{2}\leqslant t<\frac{\log(1/p)}{2}
\end{cases}
}_{\psi(t,p)},
\end{align}
and observe that the following numerical inequality holds
\newcommand{\tupper}{\frac{\log(1/2p)}{2}}
\begin{align}\label{eq:poisson_majorization}
    \psi(t,p)\leqslant \frac{\mathrm{e}^{Kt}-\frac{K^2t^2 }{2}-Kt-1}{K^2/2},\quad 0<t\leqslant \tupper,
\end{align}
for some sufficiently large constant $K$, which gives us the bound
\begin{align}
    \mathbf{E}\mathrm{e}^{tZ}\leqslant 1+2p^2\cdot \frac{\mathrm{e}^{Kt}-Kt-1}{K^2},\quad 0<t\leqslant \tupper.
\end{align}
This specific function is chosen as the upper bound because it appears in the exponential moments of Poisson random variables, making computations tracktable.
\todoi{relation to Poisson - explain here or in intro?}
%Intuitively, the claim is true as the left-hand side is of order $O(t^3)$ when $t$ is close to zero, and $O(\mathrm{e}^{6t})$ when $t$ is far from zero; in turn the upper-bounding function is of order $O(Kt^3)$ when $t$ is close to zero and $O(\mathrm{e}^{tK}/K^2)$ when $t$ is far from zero.

We know prove that \Cref{eq:poisson_majorization} holds with $K=50$; to this end we use the series expansion and compare the coefficients. Specificailly, we consider the two cases:
\begin{enumerate}
    \item For $0<t<\frac{1}{2}$ the LHS is bounded by 
$6 t^3+\sum_{r\geqslant 3}\frac{4^{r}t^r}{r!}$ (for $p\leqslant \frac{1}{30},t\leqslant \frac{1}{2}$ we have $\parta \leqslant 6t^3$) while the RHS equals $\sum_{r\geqslant 3}\frac{2K^{r-2}t^r}{r!}$; we find that the coefficients are accordingly dominated when $4^{r}\leqslant 2K^{r-2}$ for $r\geqslant 4$ and when 
$\frac{64}{3!}+6\leqslant \frac{2K}{3!}$. These conditions are satisfied for $K=50$.
    \item For $\frac{1}{2}\leqslant t\leqslant \tupper$ the LHS is bounded by $1+4t+8t^2+2\sum_{r\geqslant 3}\frac{4^r t^r}{r!}$ while the RHS equals $\sum_{r\geqslant 3}\frac{2K^{r-2}t^r}{r!}$. It suffices when
    $4^{r}<2 K^{r-2}$ for $r\geqslant 4$ and $1+4t+8t^2+\frac{2t^3}{3!}\leqslant \frac{2 K t^3}{3!}$. These conditions are satisfied when $K=50$ and $\frac{1}{2}\leqslant t$.
\end{enumerate}

\subsection{Aggregating Bounds}
Applying the last bound to $Z_i\sim^{iid} Z$ and using the inequality $\log(1+u)\leqslant u$, we get
\begin{align}
\log\mathbf{E}\mathrm{e}^{t\cdot(Z_1+\cdots+Z_m)}\leqslant 2mp^2\cdot \frac{\mathrm{e}^{Kt} - Kt -1 }{K^2},\quad 0<t<\tupper.
\end{align}
Using the Cramer-Chernoff inequality with $t^{*}=\frac{\log(1+K u /2mp^2)}{K}$ we obtain (note: calculations may be facilitated with symbolic algebra software (see \Cref{app:chernoff-subpoiss}))
\begin{align}
   \mathbf{P}\left\{Z_1+\cdots+Z_m\geqslant u\right\}\leqslant \mathrm{e}^{-\frac{u^2}{4mp^2}\cdot h\left(\frac{Ku}{2mp^2}\right)},\quad h(u)\triangleq \frac{(1+u)\log(1+u)-u}{u^2/2},
\end{align}
for all $u$ satisfying the condition
\begin{align}
    \ \frac{\log(1+K u/2mp^2)}{K}\leqslant \tupper.
\end{align}
For our case we use $u=s\epsilon$, then the condition becomes $\frac{\log(1+K\epsilon/2p)}{K}\leqslant \tupper$, and is satisfied when $\epsilon < p\log(1/2p)$ (we use $\log(1+u)\leqslant u$ with $u=K\epsilon/2p$). We then obtain
\begin{align}
   \mathbf{P}\left\{\frac{Z_1+\cdots+Z_m}{s}\geqslant \epsilon\right\} \leqslant \mathrm{e}^{-\frac{m\epsilon^2}{4}\cdot h\left(\frac{K\epsilon}{2p}\right)}.
\end{align}
Due to \Cref{eq:MGF_bound}, this discussion implies
\begin{align}
 \max\{\mathbf{P}\left\{x^T B x \geqslant \epsilon\right\}, \mathbf{P}\left\{x^T B x \leqslant -\epsilon\right\}\} \leqslant \mathrm{e}^{-\frac{m\epsilon^2}{4}\cdot h\left(\frac{K\epsilon}{2p}\right)},
\end{align}
which completes the proof.

\bibliography{citations}

\appendix

\section{Chernoff Bounds for Sub-Poisson Tails}\label{app:chernoff-subpoiss}

Suppose that $\mathbf{E}[\mathbf{e}^{tX}]\leqslant \mathrm{e}^{v^2\cdot \frac{\mathrm{e}^{Kt}-Kt-1}{K^2}}$, we then show that the Chernoff inequality gives $\mathbf{P}\{X>\epsilon\}\leqslant \mathrm{e}^{- \frac{\epsilon^{2}}{2 v} \cdot  2 \left(\frac{K \epsilon}{v}\right)^{-2} \left(\left(1 + \frac{K \epsilon}{v}\right) \log{\left(1 + \frac{K \epsilon}{v} \right)} - \frac{K \epsilon}{v}\right)}$. See the code for \Cref{alg:chernoff_sub_poiss}.
\begin{lstlisting}[language=Python,basicstyle=\footnotesize,showstringspaces=False,caption={Chernoff Inequality},label={alg:chernoff_sub_poiss}]
import sympy as sm

sm.init_printing(use_latex=True)

v,K,t,eps,u = sm.symbols('v K t epsilon u')

log_MGF = v*(sm.exp(K*t)-K*t-1)/K**2
t_best = sm.solve((log_MGF-t*eps).diff(t),t)[0]

logp = (log_MGF-t*eps).subs({t:t_best}).factor()

out = logp.subs({K*eps:u*v}).factor().collect(sm.log(u+1))
bennet = ((1+u)*sm.log(1+u)-u)/(u**2/sm.Rational(2,1))
gauss = (out/bennet)
gauss = gauss.subs({u:K*eps/v}).simplify()
bennet = bennet.subs({u:sm.UnevaluatedExpr(K*eps/v)})

gauss*bennet
\end{lstlisting}
\end{document}